\documentclass[3p,times,twocolumn]{elsarticle}

\usepackage{ecrc}

\volume{00}

\firstpage{1}

\journalname{Nuclear and Particle Physics Proceedings}

\runauth{}

\jid{nppp}

\jnltitlelogo{Nuclear and Particle Physics Proceedings}

\usepackage{amssymb}

\usepackage[figuresright]{rotating}

\newcommand\fonll{\textsc{fonll}}

\newcommand\vusphydro{v-\textsc{usp}hydro}

\newcommand\ROOT{\textsc{Root}}
\newcommand\pythia{\textsc{Pythia8}}

\newcommand\pt{\ensuremath{p_T}}
\newcommand\vnn{\ensuremath{v_n}}
\newcommand\vn[1]{\ensuremath{v_{#1}}}

\newcommand\cum[1]{\ensuremath{\{#1\}}}

\newcommand\DD{D$^0$}
\newcommand\BB{B$^0$}

\begin{document}

\begin{frontmatter}



\dochead{}

\title{Heavy meson flow harmonics in event-by-event viscous relativistic hydrodynamics}

 \author[a]{Caio A.~G.~Prado}
 \author[b]{Jacquelyn Noronha-Hostler}
 \author[a]{Jorge Noronha}
 \author[a]{Alexandre A.~P.~Suaide}
\author[a]{Marcelo G.~Munhoz}
\author[c]{Mauro R.~Cosentino}

 \address[a]{Instituto de F\'{i}sica, Universidade de S\~{a}o Paulo, C.P. 66318, 05315-970 S\~{a}o Paulo, SP, Brazil}
 \address[b]{Department of Physics, University of Houston, Houston TX 77204, USA}
 \address[c]{Centro de Ci\^{e}ncias Naturais e Humanas, Universidade Federal do ABC, Av.~dos Estados 5001, Bairro Santa Terezina, 09210-508 Santo Andr\'{e}, SP, Brazil}

\begin{abstract}
Event-by-event viscous hydrodynamics is combined with heavy quark energy loss models to compute heavy flavor flow cumulants $\vn2\cum2$, $\vn3\cum2$, and $\vn2\cum4$ as well as the nuclear modification factors of $D^0$ and $B^0$ mesons in PbPb collisions at 2.76 TeV. Our results indicate that bottom quarks can flow as much as charm quarks in the $p_T$ range 8--30 GeV.  

\end{abstract}


\end{frontmatter}


\section{Introduction}
\label{intro}

Due to the (nearly) perfect fluidity property of the Quark-Gluon Plasma (QGP) \cite{Gyulassy:2004zy}, spatial gradients present in the early stages of heavy ion collisions are efficiently converted into final state flow anisotropies and the soft limit ($p_T<2 $ GeV) of the anisotropic flow coefficients, $v_n$, can be very accurately described using event-by-event viscous hydrodynamics (see, for instance, \cite{Bernhard:2016tnd,Noronha-Hostler:2015uye}).

On the other hand, beyond the soft limit the underlying physical mechanism behind anisotropic flow changes and, at high $p_T$, the differences in the path length of jets plowing through the plasma are expected \cite{Wang:2000fq,Gyulassy:2000gk} to drive the observed $v_n$'s. This picture has been recently confirmed by event-by-event jet energy loss + viscous hydrodynamics calculations \cite{Noronha-Hostler:2016eow,Betz:2016ayq,Noronha:2016yhf,Noronha-Hostler:2016opk}, which showed that there is an approximate linear response relation (valid event-by-event) between the high $p_T$ \vn2 and the initial state (energy density) eccentricity $\varepsilon_2$. 

Motivated by the large azimuthal anisotropy displayed by open heavy flavor mesons~\cite{Abelev:2013lca,Abelev:2014ipa,Adam:2016ssk}, in this proceedings we investigate how spatial fluctuations in the initial state, and their subsequent bulk evolution via event-by-event viscous hydrodynamics, affect heavy flavor flow harmonics with $p_T \gtrsim 10$ GeV. This is done by combining a heavy quark energy loss model
with event-by-event viscous hydrodynamic backgrounds, which allow for computing 
the nuclear modification factor, $R_{AA}$, of $D^0$ and $B^0$ mesons and their corresponding flow coefficients $\vn2\cum2$, $\vn3\cum2$, and $\vn2\cum4$. 

\section{Event-by-event hydrodynamics + heavy quark energy loss model}
\label{ebehydrojet}

In order to study the evolution of heavy quarks in the medium a new
modular Monte Carlo code in C++ was developed, using \ROOT~\cite{Brun:1997pa} and
\pythia~\cite{Sjostrand:2007gs} libraries, which can be used to implement different types of energy loss models. Bottom and charm quarks in the medium are sampled with
initial momentum distribution given by \fonll\,
calculations~\cite{Cacciari:1998it,Cacciari:2001td}. The heavy quark produced at a given point $\vec{x}_0$ in the transverse plane and it moves (on the plane) with velocity $v$ along the (constant) direction
defined by an angle $\varphi_{\rm quark}$. We use a simple parametrization for the heavy quark energy loss per unit length
\begin{equation}
\label{eq:egyloss}
  \frac{dE}{dx}(T,v) = f(T,v)\,\Gamma_{\rm flow},
\end{equation}
where $T$ is the local temperature, $\Gamma_{\rm flow}
=\gamma\left[1-v\cos(\varphi_{\rm quark}-\varphi_{\rm flow})\right]$ (with
$\gamma = 1/\sqrt{1-v^2}$) \cite{Baier:2006pt}, and
$\varphi_{\rm flow}$ is the local azimuthal angle of the underlying flow.
We consider two simple models for each quark flavor: $f(T,v)  = \xi T^2$ and $f(T,v) = \alpha$. The $\sim T^2$ expression is motivated by conformal AdS/CFT
calculations \cite{Gubser:2006bz} while the temperature independent expression was chosen after the study performed in Ref.\ \cite{Das:2015ana} (other more realistic expressions for heavy flavor energy loss are currently being implemented). The free parameters $\xi$ and $\alpha$ in the energy loss expressions are fixed by matching our 
calculations for $D^0$ $R_{AA}$ to data at $p_T \sim 10$ GeV in the 0--10\% centrality class while the parameters for the bottom energy loss are found via matching our non-photonic electron $R_{AA}(p_T \sim 10 \,{\rm GeV})$ to data in the same centrality class (we use FONLL spectra for electron
production from a given quark).  

The temperature and flow profiles used in the heavy flavor calculations are
generated using the event-by-event relativistic
viscous hydrodynamical model,
\vusphydro~\cite{Noronha-Hostler:2013gga,Noronha-Hostler:2014dqa,Noronha-Hostler:2015coa},
a boost invariant code whose accuracy has been checked against known analytical solutions \cite{Marrochio:2013wla}. 
We use MCKLN initial conditions for PbPb
$\sqrt{s}=2.76$~TeV collisions \cite{Drescher:2006pi,Drescher:2007ax,Drescher:2006ca},
$\eta/s=0.11$, and an initial time $\tau_0$=0.6 fm, which leads to a good description of experimental data for the flow harmonics at low $p_T$. At the moment, coalescence is not yet implemented and hadronization in the heavy flavor sector is assumed to occur when the local temperature falls below a certain value $T_d$ at which fragmentation \cite{Peterson:1982ak} is performed. Also, we note that the energy and momentum lost by the jets are assumed to not affect the underlying bulk evolution of the medium. The effects of jet energy loss on low $p_T$ flow harmonic coefficients computed event by event were considered in \cite{Andrade:2014swa}.

Our event-by-event analysis (consisting of a couple thousand hydro events in each centrality
class) is performed by oversampling each
hydro event with millions of heavy quarks, which allows us to compute the nuclear modification factor $R_{AA}^Q(\pt,\varphi)$ and its corresponding $\vnn^Q(\pt)$ (with $Q=c,b$ being the quark flavors) flow coefficients \cite{Poskanzer:1998yz}. Clearly, very few heavy quarks are produced per event and, thus, over oversampling is meant to give the probability to find $\vnn^Q(\pt)$ in a hydro event with a certain \vnn in the soft sector. Multi-particle cumulants \cite{Luzum:2012da,Luzum:2013yya} are computed at high $p_T$ following the detailed analysis performed in Refs.\ \cite{Betz:2016ayq,Gardim:2016nrr}.

\section{Results}
\label{results}

\begin{figure} [h]
  \centering
  \includegraphics[width=0.4\textwidth]{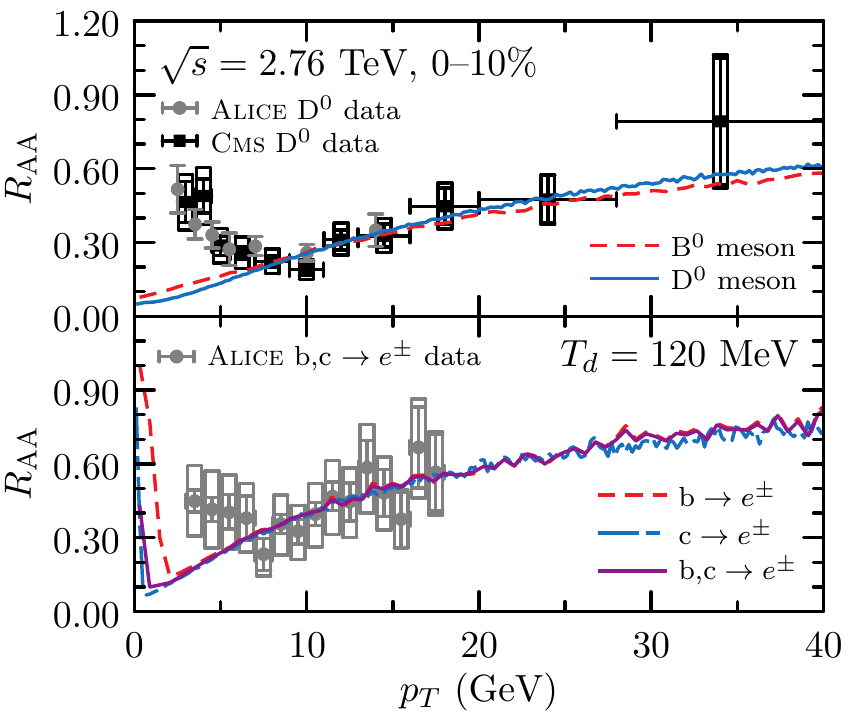}
  \caption{(Color online) Model calculations for $D^0$ $R_{AA}$ (upper panel) and heavy flavor
  electron (lower panel) using $\frac{dE}{dx} \sim \alpha$ and $T_d=120$~MeV. Experimental data from Refs.\ \cite{ALICE:2012ab,Adam:2015sza,CMS:2015hca,MoreiradeGodoy:2014vpa}.}
  \label{fig:raaBoth}
\end{figure}

We show in Fig.~\ref{fig:raaBoth} a comparison of our calculations for $D^0$ meson and heavy flavor
electron $R_{AA}$, performed using $\frac{dE}{dx}
\sim \alpha$ and $T_d=120$~MeV, with
data \cite{ALICE:2012ab,Adam:2015sza,CMS:2015hca,MoreiradeGodoy:2014vpa}. Predictions for the $R_{AA}$ of $B^0$ mesons can also be found in \ref{fig:raaBoth}. One can see that the heavy meson $R_{AA}$'s are found to be nearly the same in the $p_T$ range considered (this also holds for the other energy loss model).  

\begin{figure}[h]
  \centering
  \includegraphics[width=0.4\textwidth]{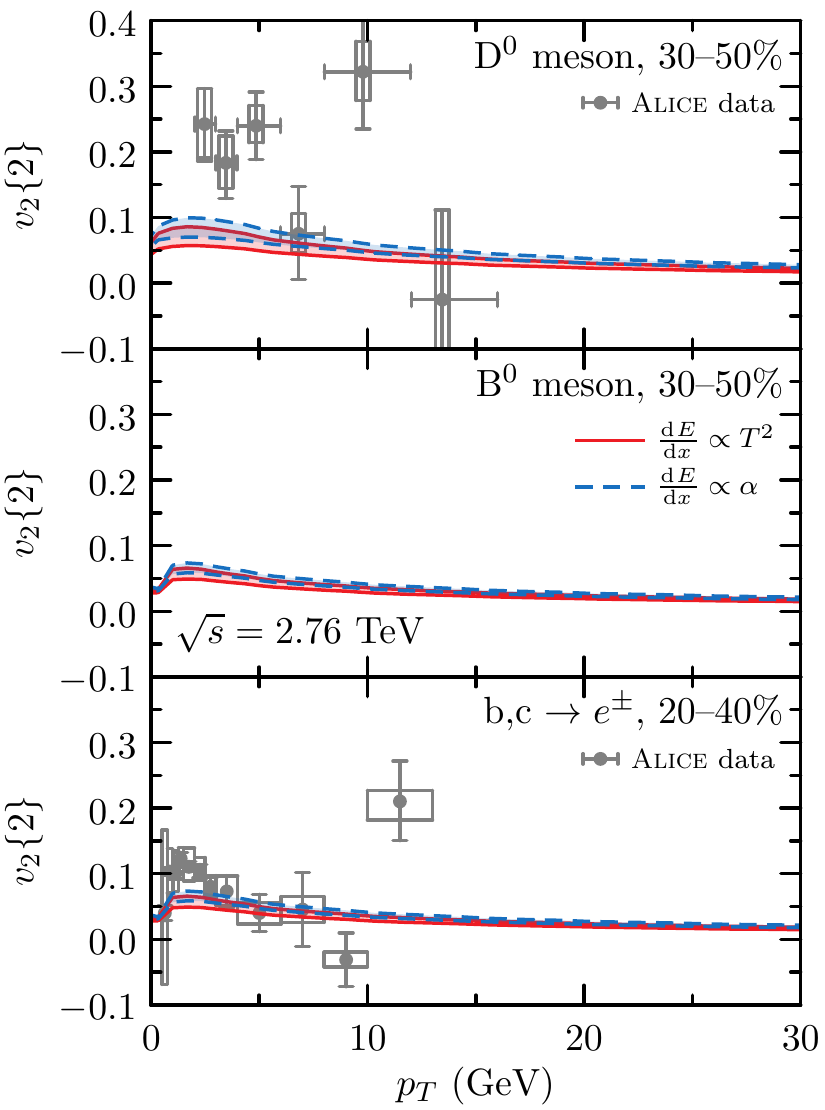}
  \caption{(Color online) Model calculations for the $\vn2\cum2$ of \DD meson (upper panel), \BB meson
  (middle), and electron from heavy flavor (lower panel) for our two energy loss
  models. The bands denote the variation in $T_d$ in the range 120--160~MeV.
  The experimental data shown are from Refs.\ \cite{Abelev:2014ipa,Abelev:2013lca,Adam:2016ssk}.}
  \label{fig:v2}
\end{figure}

The upper panel of Fig.~\ref{fig:v2} shows our results for $D^0$ meson $\vn2\cum2$ in the 30--50\% centrality class compared to experimental data while in the lower panel a comparison between our calculations and electron data in the 20--40\% centrality class is shown. One can see that our calculations are consistent with the data at high $p_T$. However, at low $p_T$ where coalescence effects are not negligible our results fall below the experimental data. In this low $p_T$ regime, other effects concerning the heavy quark energy loss mechanism (not taken into account here) become increasingly important \cite{Nahrgang:2014vza}, which can affect the overall magnitude of the computed flow harmonics. 

The middle panel of Fig.~\ref{fig:v2} shows our corresponding results for $B^0$ meson $\vn2\cum2$ at $\sqrt{s}=2.76$~TeV in the 30--50\% centrality class. One can see that the approximate ``flavor independence" previously found in the meson $R_{AA}$'s displayed in Fig.~\ref{fig:raaBoth} is carried over to the flow harmonics. Thus, in our model bottom quarks flow just as much as charm quarks do in the $p_T$ range 10--30 GeV and the fact that the rest mass of bottom quarks is a factor of $\sim 3$ larger than the charm quark mass does not seem to matter for the final flow harmonics in this case. 

While the energy loss models lead to similar results, we did observe that the case in which $\frac{dE}{dx} \sim \alpha$ produces the largest elliptic flow, which gives support to the analysis done in \cite{Das:2015ana} using an averaged initial condition. Furthermore, we find that $\vn2\cum2$ increases with decreasing $T_d$, which is similar to what happens in the soft sector (in that case, the low $p_T$ $\vn2\cum2$ would increase with decreasing freeze-out temperature). This shows that remaining uncertainties in the hadronization mechanism can affect the magnitude of the elliptic flow in our model.

\begin{figure}[h]
  \centering
  \includegraphics[width=0.4\textwidth]{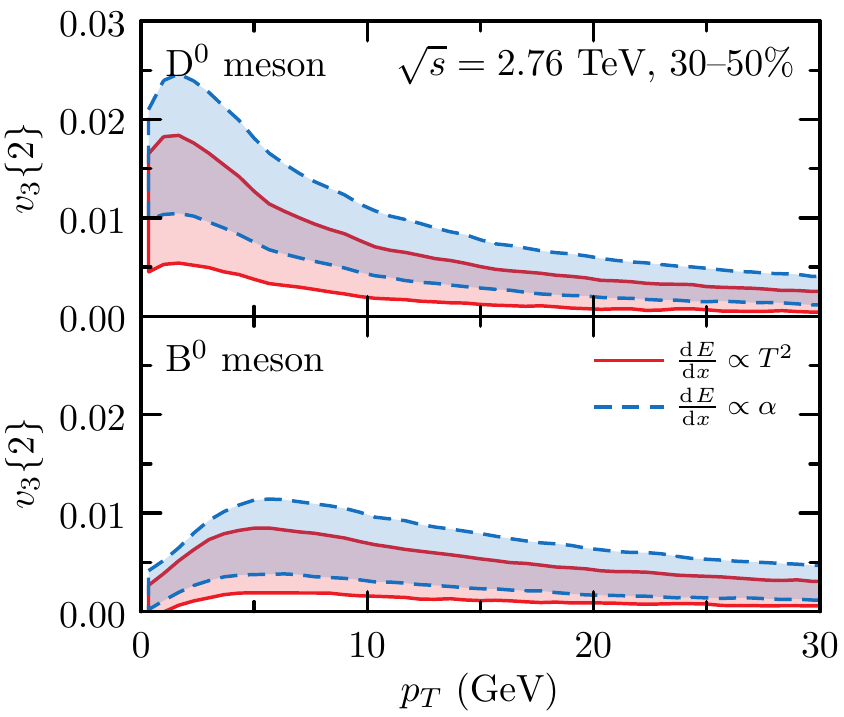}
  \caption{(Color online) Model calculations for the $\vn3\cum2$ of \DD meson (upper panel) and \BB
  meson (lower panel) for two energy loss models. The bands represent the
  variation of the $T_d$ range 120--160~MeV.}
  \label{fig:v3}
\end{figure}

With the advent of event-by-event calculations, one can also compute $\vn3\cum2$ of $D^0$ and $B^0$ mesons and our results for these quantities are shown in Fig.~\ref{fig:v3}. One can see that the different masses of the heavy quarks do not affect the final triangular flow in the $p_T$ range considered since $\vn3\cum2$ is basically the same for $D^0$ and $B^0$ mesons. However, there are a couple of interesting points worth noticing about our results for the triangular flow of heavy mesons. First, our $\vn3\cum2$ is found to be approximately an order of magnitude smaller than $\vn2\cum2$. We note that the first calculations of triangular flow of $D^0$ mesons (within the event-plane method) were done in Refs.\ \cite{Nahrgang:2014vza,Nahrgang:2016lst}, and a comparison between our results shows that our $\vn3\cum2$ is smaller than the triangular flow result they reported. Also, we found that $\vn3\cum2$ is much more sensitive to how the heavy quarks decouple from the medium, i.e., this quantity depends more strongly on the hadronization parameter $T_d$. 

\begin{figure}[h]
  \centering
  \includegraphics[width=0.4\textwidth]{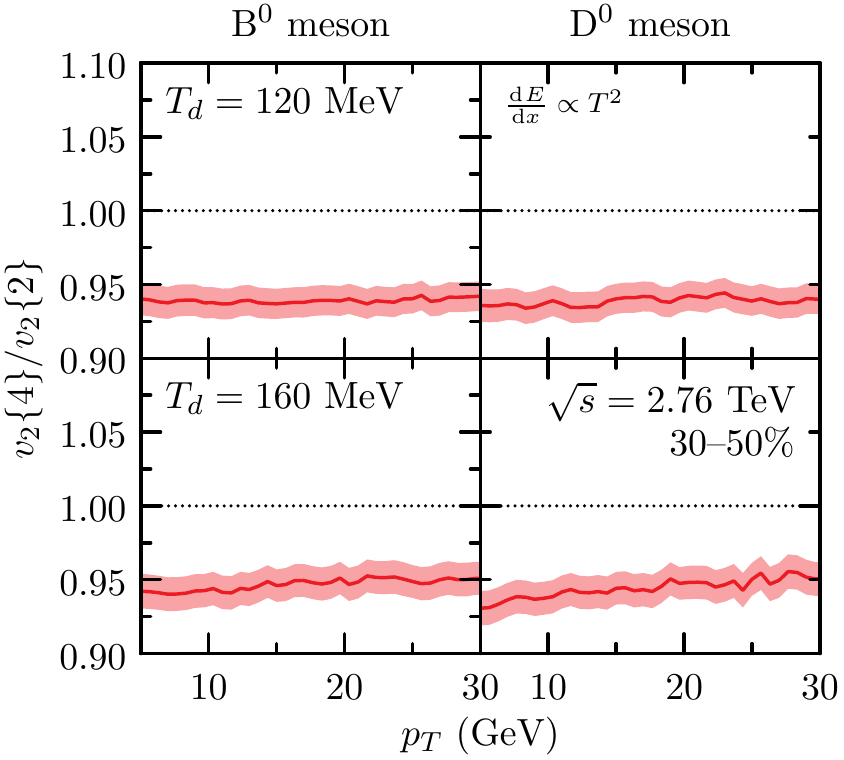}
  \caption{(Color online) Model calculations for the ratio $\vn2\cum4/\vn2\cum2$ as a function of $p_T$ for $D^0$ and $B^0$ mesons using $dE/dx \sim T^2$ and $T_d = 120$ MeV and $160$ MeV. The bands represent the systematic errors.}
  \label{fig:cumulant}
\end{figure}

In order to obtain a better understanding of how the fluctuations in the soft sector affect the anisotropic flow of heavy flavor, we follow the discussion done in \cite{Betz:2016ayq} and compute the differential elliptic flow 4-particle cumulant $\vn2\cum4$ defined by the correlation of three soft particles with one in the heavy flavor sector. The ratio between  $\vn2\cum4/\vn2\cum2$ as a function of $p_T$ for $D^0$ and $B^0$ mesons in the 30-50\% centrality class is shown in Fig.\ \ref{fig:cumulant}. One can see that this ratio is nearly constant in $p_T$ being very close to unity for both $D^0$ and $B^0$ mesons (regardless of the energy loss model and the value of $T_d$). A similar result was found for this observable in the light flavor sector \cite{Betz:2016ayq}, which suggests a common origin for the fluctuations (i.e., the initial state inhomogeneities). 

\section{Conclusions}
\label{conclusions}

By combining event-by-event hydrodynamic flow and temperature profiles with heavy quark energy loss calculations one is able to describe $R_{AA}$ and \vn2 of $D^0$ mesons and heavy flavor
electrons in the $p_T$ range 8--30~GeV of $\sqrt{s}=2.76$ TeV PbPb collisions at LHC. This type of modeling was used to obtain predictions for the corresponding $\vn2\cum2$ and $\vn3\cum2$ of $B^0$ mesons at the same collision energy. In our model $\vn2\cum2$ of $D^0$ and $B^0$ mesons are nearly the same in
this $p_T$ range, a feature that could be experimentally verified in the near future (most likely using LHC run 2 data at $\sqrt{s}=5.02$ TeV). Once initial state fluctuations, and their evolution using event-by-event viscous hydrodynamics, are taken into account new observables, such as the 4-particle cumulant $\vn2\cum4$, can be computed to investigate how fluctuations in the soft sector may affect the anisotropic flow of heavy flavor. 

It is evident from Figs.\ \ref{fig:raaBoth}-\ref{fig:cumulant}, however, that these observables are not yet sensitive enough to capture the differences between the two energy loss models used in this work. By extending event engineering techniques \cite{Schukraft:2012ah,Aad:2015lwa} to the heavy flavor sector new observables that are more sensitive to the underlying energy loss model may be studied. The first study in this direction was performed in \cite{Prado:2016szr} where the correlation between the soft hadron $v_2$ and the heavy meson $v_2$ was computed, which revealed an interesting linear correlation between these observables on an event-by-event basis.

\section*{Acknowledgements}

The authors thank M.~Luzum for discussions about flow cumulants and Funda\c{c}\~ao de Amparo \`a Pesquisa do Estado de S\~ao Paulo (\textsc{fapesp}) and Conselho Nacional de Desenvolvimento Cient\'ifico
e Tecnol\'ogico (\textsc{cnp}q) for support. J.N.H. was supported by the
National Science Foundation under grant no. PHY-1513864 and she acknowledges
the use of the Maxwell Cluster and the advanced support from the Center of
Advanced Computing and Data Systems at the University of Houston to carry out
the research presented here. J.N. thanks the University of Houston for its
hospitality.

\bibliographystyle{elsarticle-num}
\bibliography{library}







\end{document}